\documentclass[aps,prl,superscriptaddress,twocolumn,longbibliography,floatfix]{revtex4-2}
\usepackage{units}
\usepackage{amsmath}
\usepackage{amsthm}
\usepackage{amssymb}
\usepackage{graphicx}
\usepackage[english]{babel}
\usepackage[utf8]{inputenc}
\usepackage{mathtools}
\usepackage{color}
\usepackage{xcolor}
\usepackage{bbold}
\usepackage{braket}
\usepackage{tabularray}
\usepackage[version=3]{mhchem}

\definecolor{myurlcolor}{rgb}{0,0,0.7}
\definecolor{myrefcolor}{rgb}{0.8,0,0}

\usepackage[
	breaklinks,
	pdftex,
	colorlinks=true, 
	linkcolor=myrefcolor,
	citecolor=myurlcolor,
	urlcolor=myurlcolor
]{hyperref}
\usepackage{cleveref}

\usepackage[linesnumbered, ruled,vlined]{algorithm2e}

\graphicspath{{./images/},{./imagesAppendix/},{./Plots/}}

\renewcommand{\eqref}[1]{Eq.~(\ref{#1})} 

\def\app#1#2{%
  \mathrel{%
    \setbox0=\hbox{$#1\sim$}%
    \setbox2=\hbox{%
      \rlap{\hbox{$#1\propto$}}%
      \lower1.1\ht0\box0%
    }%
    \raise0.25\ht2\box2%
  }%
}

\ifx\proof\undefined
\newenvironment{proof}[1][\protect\proofname]{\par
	\normalfont\topsep6\p@\@plus6\p@\relax
	\trivlist
	\itemindent\parindent
	\item[\hskip\labelsep\scshape #1]\ignorespaces
}{%
	\endtrivlist\@endpefalse
}
\providecommand{\proofname}{Proof}
\fi
\makeatother

\newcommand{\idg}[1]{{\bfseries #1)}}

\providecommand{\ftname}{ft}
\providecommand{\theoremname}{Theorem}
\providecommand{\claimname}{Claim}
\providecommand{\lemmaname}{Lemma}
\providecommand{\definitionname}{Definition}

\definecolor{KB}{rgb}{0.4,0.3,0.9}

\definecolor{THc}{rgb}{0.9,0.3,0.2}

\newcommand{\prlsection}[1]{{\em {#1}.---~}}

\newcommand{\sectionMain}[1]{
\let\oldaddcontentsline\addcontentsline
\renewcommand{\addcontentsline}[3]{}
\section{#1}
\let\addcontentsline\oldaddcontentsline
}

\allowdisplaybreaks

\begin{document}

\title{Rydberg atomtronic devices}

\author{Philip Kitson}
\affiliation{Quantum Research Center, Technology Innovation Institute, P.O. Box 9639 Abu Dhabi, UAE}
\affiliation{Dipartimento di Fisica e Astronomia ``Ettore Majorana'', Via S. Sofia 64, 95123 Catania, Italy}
\affiliation{INFN-Sezione di Catania, Via S. Sofia 64, 95123 Catania, Italy}
\author{Tobias Haug}
\affiliation{Quantum Research Center, Technology Innovation Institute, P.O. Box 9639 Abu Dhabi, UAE}
\author{Antonino La Magna}
\affiliation{National Research Council, Institute for Microelectronics and Microsystems (IMM-CNR), VIII Strada 5, Catania, 95121, Italy}
\author{Oliver Morsch}
\affiliation{CNR-INO and Dipartimento di Fisica dell’Universit\`{a} di Pisa, Largo Pontecorvo 3, 56127 Pisa, Italy}
\author{Luigi Amico}
\affiliation{Quantum Research Center, Technology Innovation Institute, P.O. Box 9639 Abu Dhabi, UAE}
\affiliation{Dipartimento di Fisica e Astronomia ``Ettore Majorana'', Via S. Sofia 64, 95123 Catania, Italy}
\affiliation{INFN-Sezione di Catania, Via S. Sofia 64, 95123 Catania, Italy}
\affiliation{Centre for Quantum Technologies, National University of Singapore 117543, Singapore}

\begin{abstract}
    Networks of Rydberg atoms provide a powerful basis for quantum simulators and quantum technologies. Inspired by matter-wave atomtronics, here we engineer switches, diodes and universal logic gates composed of Rydberg networks. Our schemes control the dynamics of a Rydberg excitation via the anti-blockade or facilitation mechanism, allowing for much faster devices compared to cold atom systems. Our approach is robust to noise and can be applied to individually trapped atoms and extensive three-dimensional gases. In analogy to electronics, Rydberg atomtronic devices promise to enhance quantum information processors and quantum simulators.
\end{abstract}

\maketitle

\prlsection{Introduction} Electrons in Rydberg atoms can be excited to very large principle quantum number~\cite{browaeys2020many,adams2019rydberg,morsch2018dissipative}. The resulting large dipole moment and polarisability lead to peculiar effects, such as the dipole blockade: within a specific volume, the excitation of more than one atom to the Rydberg state is inhibited due to the aforementioned dipole interaction~\cite{comparat2010dipole}. Conversely, when the excitation laser is negatively detuned from resonance, an anti-blockade or facilitation effect occurs: a single initial excitation induces more excitations in neighbouring atoms~\cite{valado2016experimental}. Combining blockade and facilitation effects together can provide flexible schemes for coherent manipulation of excitations in networks of Rydberg atoms~\cite{leonhardt2014switching,gutierrez2017experimental}. The inherent physics and the remarkable know-how in coherent atom manipulations~\cite{endres2016atom, barredo2016atom, schymik2020enhanced}, networks of Rydberg atoms provide a fruitful and versatile toolbox for quantum simulators and more widely quantum technologies~\cite{browaeys2020many,morsch2018dissipative,bluvstein2021controlling, bernien2017probing, barredo2015coherent, chen2023continuous,morgado2021quantum,henriet2020quantum,adams2019rydberg,wu2021concise}. Rydberg networks also provide a promising basis for quantum information processors~\cite{saffman2010quantum, cong2022hardware, lukin2001dipole}.

In our approach, we are inspired by atomtronics, which encapsulates the properties of ultra-cold atoms to create circuits via laser fields of different shapes and intensities~\cite{seaman2007atomtronics, amico2005quantum, amico2022colloquium,amico2021roadmap,polo2024perspective}. In particular, atomic devices such as atomtronic transistors and switches for cold atoms have been proposed~\cite{gajdacz2014atomtronics, stickney2007transistorlike,wilsmann2018control} and realised~\cite{caliga2016transport}. Another vital building block to perform classical analogue or digital computation is the diode. In the same way as electronics, the atomtronic diode has been proposed by bringing doped conducting cold atom systems together~\cite{pepino2009atomtronic, pepino2010open,seaman2007atomtronics}. 

Here, we demonstrate how the aforementioned control of Rydberg excitations can be exploited to conceive specific atomtronic devices in which, instead of matter, the dynamics involve Rydberg excitations. The transfer and control of excitations are conducted via the facilitation mechanism, where an excited state of an atom induces excitations in neighbouring atoms via the van der Waals interaction combined with appropriately chosen frequency detunings. By applying this idea to different networks, we construct specific Rydberg atomtronic schemes analogous to switches and diodes. Further, we construct logic gates such as AND, NOT and NAND, demonstrating that Rydberg atomtronics provides a universal logic gate set. 
\begin{figure}[htbp]
    \centering
\includegraphics[width=0.3\textwidth]{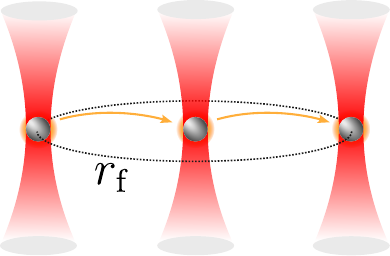}
    \caption{Network of Rydberg atoms trapped with optical tweezers. To create Rydberg atomtronic devices, we control the flow of excitations between atoms using the facilitation mechanism: An excitation in the first atom induces an excitation in a neighbouring atom at facilitation distance $r_{\text{f}}$ only for a facilitation detuning $\Delta_{\text{f}}=-C_6/r_{\text{f}}^6$, else transport is suppressed.}
    \label{Network_Ryd_Atoms}
\end{figure}

\vspace{-2mm}
A key component for these devices, especially for the diode, is the generation of a non-reciprocal or chiral flow of excitations. When considering interactions within two levels of different Rydberg excitations, chiral currents in ring-shaped networks have been induced via phase shifts~\cite{perciavalle2023controlled}. In contrast, here we consider the dynamics of the ground state and excited Rydberg state which lacks a coherent hopping interaction. Nonetheless, we can engineer non-reciprocal behaviour by spatially varying the distance and detuning of atoms to create a one-way facilitation mechanism.

\prlsection{Model}
We study the dynamics of excitations in a network of $N$ Rydberg atoms where we denote the atomic ground state as 
$\ket{\downarrow}$ and Rydberg state as $\ket{\uparrow}$ with Hamiltonian~\cite{morsch2018dissipative}, 
\begin{equation}
    \mathcal{H} = \sum_{j=1}^N \, \Delta_{j} \, n_{j} \, + \, \Omega \, \sum_{j=1}^N \, \sigma^{x}_j \, + \, \frac{1}{2} \, \sum_{i\ne j} \, \frac{C_{6}}{{|x_{i} - x_{j}|}^{{6}}} \, n_{i}n_{j}.
    \label{Hamiltonian}
\end{equation} 
Here, $\Omega$ is the Rabi frequency, $\Delta_{j}$ the detuning of the $j^{\text{th}}$ atom for the $x$-Pauli $\sigma^{x}_j$, $C_6$ the van der Waals interaction coefficient,  $n_{j} = \frac{1}{2}(\sigma_{j}^{z} + \mathbb{1})$ the excitation number operator and $x_j$ the position of the $j^{\text{th}}$ atom. Coupling with the environment for mixed state $\rho$ is modelled with the Lindblad master equation,
\begin{equation}
    \partial_{t}\rho = -i[\mathcal{H}, \, \rho] \, + \, \sum_{k} \, \left(\text{L}_{k} \, \rho \, \text{L}_{k}^{\dagger} \, - \, \frac{1}{2}\{\text{L}_{k}^{\dagger}\text{L}_{k}, \, \rho\}\,\right).
    \label{Quantum_Equation}
\end{equation} 
We consider two dissipative mechanisms expressed with Lindblad operators: dephasing $\text{L}_{k, \, \text{dephasing}} = \sqrt{\gamma} \, n_{k}$ with rate $\gamma$ as well as decay of excitations $\text{L}_{k, \text{decay}} = \sqrt{\kappa} \, \sigma_{k}^{-}$ with rate $\kappa$, where $\sigma_{k}^{-}$ destroys a Rydberg excitation. In the limit of strong dephasing $\gamma\gg \Omega$, the atoms rapidly dephase into mixed states~\cite{morsch2018dissipative} and quantum coherences can be neglected. In this regime, a classical master equation can be derived via a second-order perturbation theory \cite{lesanovsky2013kinetic, marcuzzi2014effective}. The evolution of the probabilities of the basis states $\textbf{p}=\text{diag}(\rho)$ is given by
\begin{equation}
    \partial_{t}\textbf{p} = \sum_{k} \, \Gamma_{k} \, [\sigma_{k}^{+} \, + \, \sigma_{k}^{-} \, - \, 1]\textbf{p} \, + \, \sum_{k} \kappa \, [\sigma_{k}^{-} \, - \, n_{k}]\textbf{p} ,
    \label{Classical_Equation}
\end{equation}
with transition rate
\begin{equation}
    \Gamma_{k} = \frac{\Omega^{2}\gamma}{\left(\frac{\gamma}{2}\right)^{2} \, + \, \left(\Delta_{k} \, + \, \text{C}_{6} \, \sum_{q \neq k}\, \frac{n_{q}}{|x_{k} - x_{q}|^{6}}\right)^{2}} \, .
    \label{Transition_Rate_Equation}
\end{equation}
We now review two fundamental phenomena observed in Rydberg atoms. First, we illustrate the Rydberg blockade. Let us assume an atom in the ground state and detuning $\Delta=0$. When there are no other excited atoms nearby, the driving $\Omega$ will excite the atom. In contrast, if there is an excited Rydberg state within the Rydberg radius $r_\text{b}=(C_6/\Omega)^{1/6}$, the atom cannot be excited due to the energy shift induced by the van der Waals interaction.

Second, the Rydberg facilitation mechanism induces excitations only when another excitation is present~\cite{morsch2018dissipative}. Let us consider two atoms at the facilitation distance $r_{\text{f}}$ and the facilitation detuning $\Delta_{\text{f}}=-C_6/r_{\text{f}}^6$ where we choose $\vert\Delta_{\text{f}}\vert\gg \Omega$. When initially none of the atoms are excited, then the detuning will suppress any excitations due to Rabi driving $\Omega$. Now, what happens if the first atom is excited? In this case, the positive van der Waals interaction combined with the negative detuning $\Delta_{\text{f}}$ brings the second atom into resonance, as shown in Fig.~\ref{Network_Ryd_Atoms}, which induces its excitation~\cite{lesanovsky2014out}. Repeating this process, the induced excitation can further excite the next atom, effectively creating a facilitation chain of propagating excitations (see Appendix \ref{Transport_sub_section}). The facilitation mechanism is robust even for strong dephasing noise.

We now apply the blockade and facilitation mechanisms to control the flow of excitations in networks of Rydberg atoms and create various practical devices. 

\prlsection{Switch for individually trapped atoms}
A switch is a device that allows current to pass through it whilst it is enabled, however, prevents the transport of current when it is disabled. 

\begin{figure}[htbp]
    \centering
    \includegraphics[width=0.3\textwidth]{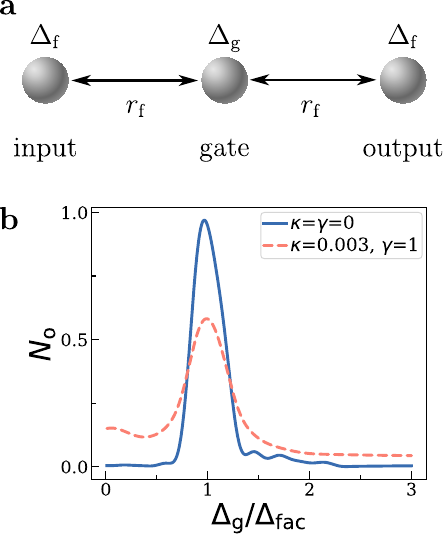}
    \caption{Switch for individually trapped atoms. Number of excitations in output $N_{\mathrm{o}}$ against gate detuning $\Delta_{\text{g}}/\Delta_{\text{f}}$ for both dissipation and non-dissipation. The used parameters are noted in Appendix \ref{Experimental_Considerations}. }
    \label{Switch_Plot}
\end{figure}

The smallest setup for an atomtronic switch is composed of a one-dimensional chain of $N = 3$ atoms with an inter-atomic distance $r_{\text{f}}$, as shown in Fig.~\ref{Switch_Plot}a. We initialise the system with an excitation in the input and all other sites in the ground state. The transport to the output is controlled by a gate atom with variable detuning $\Delta_{\text{g}}$. When we choose $\Delta_{\text{g}}\approx \Delta_{\text{f}}$, the gate atom is excited by the input via the facilitation mechanism, while otherwise, the gate atom remains with high probability in the ground state. Hence, if the gate atom is excited, the facilitation mechanism induces excitations in the output.

In Fig.~\ref{Switch_Plot}b, we vary the gate detuning $\Delta_{\text{g}}$ and measure the average number of excitations in the output $N_{\text{o}}$. We choose to record these values at t$\Omega = 3.20$ (and t$\Omega = 4.60$ when considering dephasing) corresponding to the time reached for maximum excitation in the system whilst $\Delta_{\text{g}}\hspace{-0.1cm} = \hspace{-0.1cm} \Delta_{\text{f}}$. We recognise that our device behaves like a switch, as while the gate atom is on facilitation excitations can flow to the output, however, moving away from the facilitation regime excitations cannot reach the output. The dynamics are robust against dephasing $\gamma$ and decay $\kappa$. 

\begin{figure}[htbp]
    \centering
    \includegraphics[width=0.4\textwidth]{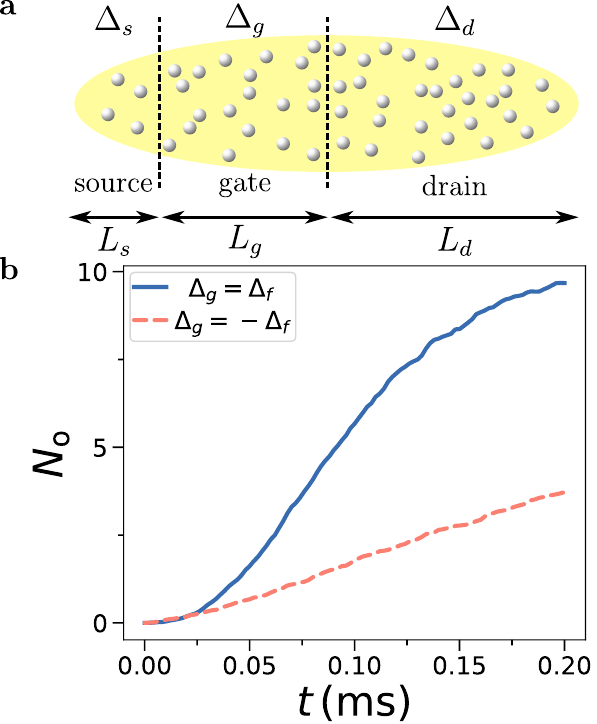}
    \caption{\idg{a} Schematic of a switch using an extensive Rydberg gas in the limit of strong dephasing~\eqref{Classical_Equation}. The gas is spilt up into three regions: the input with detuning, $\Delta_{s} = 0$, the gate with detuning either $\Delta_{g} = \Delta_{f}$ (switch on) or $\Delta_{g} = -\Delta_{f}$ (switch off) and the output with detuning $\Delta_{d} = \Delta_{f}$. For this setup, we use explicit dimensionfull parameters, which are denoted by a tilde (see Appendix \ref{Experimental_Considerations}). 
    \idg{b} Switch using $N=3000$ atoms trapped in a potential shaped as a cylinder of length $L_x=30\mu \text{m}$ and a radius of $R=7 \mu \text{m}$. We show the number of output excitations $N_{\mathrm{o}}$ controlled by gate detuning $\Delta_{g}$.
    We have $\tilde{C}_6/(2\pi) = 869 \text{GHz}\, \mu \text{m}^6$, $L_{s}=5\mu \text{m}$, $L_{g}=10\mu \text{m}$, $L_{d}=15\mu \text{m}$, $\tilde{\gamma}/(2\pi)=700 \text{kHz}$, $\tilde{\kappa}/(2\pi)=2 \text{kHz}$, $\tilde{\Omega}/(2\pi)=50 \text{kHz}$, $\tilde{\Delta_{f}}/(2\pi)=-69.5 \text{MHz}$. We have $\tilde{r_{f}}=4.8\mu \text{m}$, $\tilde{r_{b}}=10.4\mu \text{m}$ and a gas density of $2.3\cdot 10^{12} cm^{-3}$. The numerical simulation is averaged over 10 random instances of atom positions with 30 simulation trajectories each.}
    \label{Switch_3D_Gas}
\end{figure}

\prlsection{Switch for three-dimensional gas}
Next, we consider $N$ Rydberg atoms trapped in a three-dimensional potential without individual control over the position. As shown in Fig.~\ref{Switch_3D_Gas}a, we have a cylindrical trap of length $L_x$ and radius $R$, where the minimal distance between atoms is $d_{\text{min}}=0.1\mu \text{m}$. 

All atoms are subject to the same driving strength $\Omega$ and are initialised in the ground state. The system along the $x$ direction is split into three regions: input of length $L_{s}$, gate of length $L_{g}$ and output of length $L_{d}$. Each region has a different detuning frequency $\Delta(x)$: in the input, we excite Rydberg atoms on resonance with $\Delta_{s}=0$. In the gate, we have either $\Delta_{g}=\Delta_{f}$ when the switch is on, else we set $\Delta_{g}=-\Delta_{f}$ to block any transport of excitations. In the output, we set the detuning to the facilitation regime $\Delta_{d}=\Delta_{f}$. For the gate to block transport,  $L_{g}$ must be larger than the facilitation radius $r_{f}$, else excitations in the input can directly excite the output. We simulate the dynamics using the classical approximation~\eqref{Classical_Equation} by Monte-Carlo sampling of trajectories where we confirm the validity of the strong dephasing approximation in Appendix \ref{Classical_vs_Quantum_Simulation_Subsection}.

The dynamics of the excitations in the output are shown in Fig.~\ref{Switch_3D_Gas}b. Our simulation parameters are chosen closely to the ones from the Rubidium atom experiment in Ref. \cite{valado2016experimental}. For an enabled switch with $\Delta_{g}=\Delta_{f}$ we observe twice as many excitations compared to the disabled switch with $\Delta_{g}=-\Delta_{f}$. This behaviour is robust in the presence of strong dephasing and excitation loss. 

\begin{figure}[htbp]
    \centering
    \includegraphics[width=0.3\textwidth]{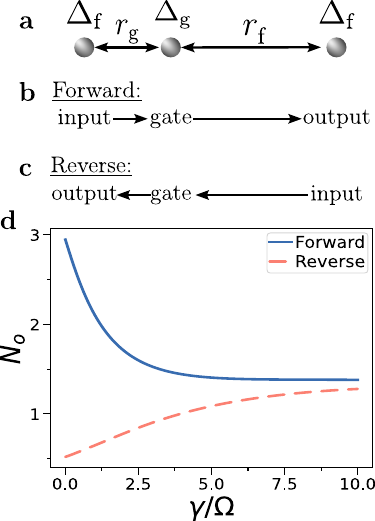}
    \caption{Atomtronic diode. \idg{a} Schematical representation of an $N = 3$ Rydberg atom diode. \idg{b} In forward operation, excitations can travel from input to output via a gate atom. Input and gate are at a distance $r_{\text{g}}$, which satisfy facilitation condition on gate atom $\Delta_{\text{g}}=-C_6/r_{\text{g}}^6$. \idg{c} In the reverse operation of the diode, excitations cannot travel from gate to output as distance $r_{\text{g}}$ does not satisfy the facilitation condition for detuning $\Delta_{\text{f}}$ on the output atom. \idg{d} We plot number of excitations $N_{\text{o}}$  of the output against dephasing $\gamma$ atoms for the forward and reverse direction with $N=6$, C$_{6} = 10$, $\kappa = 0.003$ and $\Delta_{\text{g}}/\Delta_{\text{f}}=2$ in dimensionless units}.
    \label{Diode_Plot}
\end{figure} 

\prlsection{Diode}
The diode is a non-reciprocal device that allows current to pass through one direction, but blocks transport coming from the reverse direction. We introduce non-reciprocal behaviour with the set-up in Fig.~\ref{Diode_Plot}a, where we engineer the position of the gate atom to either induce or block excitations. We define each direction in terms of the relative positions of the input and output atoms. First, in Fig.~\ref{Diode_Plot}b, we consider the forward direction of the diode. The input atom is positioned as the left-hand neighbour of the gate atom with an inter-atomic distance, $r_{\text{g}}$. In this scenario, the facilitation condition is reached, thus the initial excitation in the input can travel via the gate to the output. In contrast, we recognise the reverse direction as seen in Fig.~\ref{Diode_Plot}c, now the input is located as the right-hand neighbour to the gate at an increased distance, $r_{\text{f}}$. With the increased distance and larger detuning of the gate atom, the facilitation condition will not be met therefore blocking any transport.  

In Fig.~\ref{Diode_Plot}d, we set the detuning of the gate atom $\Delta_{\text{g}}/\Delta_{\text{f}} = 2$ (with a justification given in Appendix \ref{Diode_sub_section}) and show the number of excitations in the output $N_{\text{o}}$ at evolution time $t\Omega = 4.60$ for different values of the dephasing, $\gamma$. We find that the forward direction transports a large number of excitations compared to the reverse operation of the diode. The difference between forward and reverse decreases with increasing $\gamma$, but remains relatively large even when $\gamma$ is in the same order as the driving frequency $\Omega$. 

\prlsection{Logic gates}
We now construct different logic gates using the Rydberg interactions. Logic gates return a binary outcome depending on given input bits. For the input, we define logic $0$ as a Rydberg atom in the ground state, while $1$ corresponds to an excited Rydberg atom. We define the logic output as $0$ when $N_{\text{o}}<N_{\text{threshold}}$, while $1$ corresponds to $N_{\text{o}}>N_{\text{threshold}}$, where $N_{\text{threshold}}$ is a threshold number of excitations. 

\begin{figure}[htbp]
    \centering
    \includegraphics[width=0.45\textwidth]{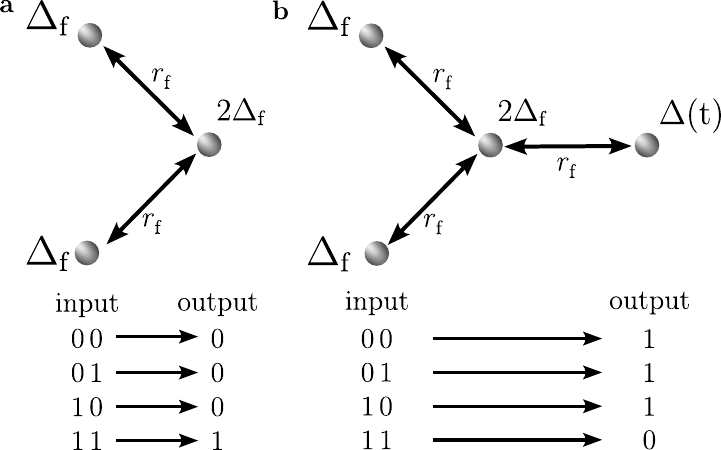}
    \caption{Logic gates with Rydberg atomtronics. \idg{a} AND gate using three Rydberg atoms. \idg{b} NAND gate, composed of four Rydberg atoms which combine an AND gate with a NOT gate.}
    \label{Gate_Schematic}
\end{figure}

The AND gate returns $1$ only when two inputs are $1$, else $0$. We construct the AND gate with three atoms as seen in Fig.~\ref{Gate_Schematic}a. The two input atoms are at a distance $r_{\text{f}}$ to the output atom, while the detuning of the output atom is chosen as $2\Delta_{\text{f}}$, i.e. twice the original facilitation condition. Only when both input atoms are excited, the output atom is on resonance due to the van der Waals interaction.

\begin{figure}[htbp]
    \centering
    \includegraphics[width=0.45\textwidth]{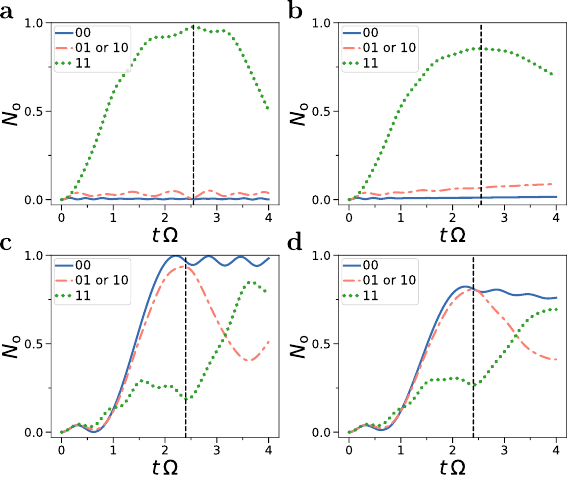}
    \caption{Number of excitation in the output of the $N =3$ AND gate (\idg{a, b}) and $N = 4$ NAND gate (\idg{c, d}). Curves show different input excitations, e.g. $01$ corresponds to the first input atom being in the ground state and the second input atom in the excited state. The dashed line indicates the time at which the gate has optimal performance. \idg{a, c} $\gamma = 0$, $\kappa = 0$. \idg{b, d} $\gamma = 1$, $\kappa = 0.003$. The parameters are: $C_{6} = 10$ and $\Delta_{\text{f}} = -\text{C}_{6}/\text{r}_{\text{f}}^{6}$ in dimensionless units. The $\pi$ pulse for the NAND gate is centered around time $t\Omega=1.5$.}
    \label{Gate_Results} 
\end{figure}

Next, we consider the NAND gate, which is an inverted AND gate, i.e. it returns $0$ only when the two inputs are $1$. We realise the NAND gate by combining the AND gate with a NOT gate (see Fig.~\ref{Gate_Schematic}b). The NOT gate flips $0$ to $1$ and vice versa. In our setup, we realise the NOT gate by setting the detuning on the output atom to $\Delta=0$ for a time period $\delta t=\pi/(2\Omega)$ centered at $t\Omega=1.5$, and $\Delta=\Delta_{\text{f}}$ for all other times. Together with the constant Rabi driving, this realises a $\pi$ pulse which excites the ground state to a Rydberg state and de-excites an initial Rydberg state into the ground state. We create a NAND gate by applying a NOT gate on an additional atom which is at facilitation distance to the output of the AND gate. 

Fig.~\ref{Gate_Results} indicates the average number of excitations $N_{\text{o}}$ in the output against time for different input excitations for both the AND gate and the NAND gate. We observe that the logic table of the AND (Fig.~\ref{Gate_Results}a,b) and NAND gate (Fig.~\ref{Gate_Results}c,d) can be realised by reading out $N_{\text{o}}$.  We find that a threshold $N_{\text{threshold}}=0.5$ is sufficient to distinguish between $0$ and $1$ even in the presence of noise. We find the optimal work time $t_{\text{w}}$ as dashed lines where we find optimal performance for the gates.

\prlsection{Discussion}
We have demonstrated that networks of Rydberg atoms can create atomtronic devices that, instead of matter-wave, are based on a  controlled flow of excitations. The flow is controlled by using the blockade and facilitation mechanism of interacting Rydberg atoms. This way, a new platform of atomtronic circuits is proposed. The propagation of matter-wave in typical cold atoms clouds occurs on the millisecond scale, whereas  Rydberg excitations can travel in microseconds. Therefore, Rydberg excitations have the potential to provide proof for fast atomtronic quantum devices. 

With this approach, we have demonstrated different circuit elements providing the Rydberg atomtronics counterpart of classical electronic devices as switches and diodes. 
In particular, diodes require non-reciprocal transport which commonly is implemented by breaking time-reversal symmetry via the flux~\cite{perciavalle2023controlled}. In contrast, we engineer non-reciprocal transport by using the facilitation mechanism combined with non-uniform atomic distances.
Further, by using the facilitation condition involving multiple atoms we implement AND, NOT and NAND classical gates, realising a universal logic gate set. Future work can combine our different gates and devices to create even more complex gadgets such as adders or routers. Additionally, collective spin excitation could also be studied within the systems. Our proposed devices use experimentally demonstrated parameter regimes and thus can be realised in state-of-the-art experiments for tweezer arrays of Rydberg atoms and three-dimensional gases.  
\prlsection{Note added} While writing the manuscript, a similar mechanism to engineer non-reciprocal transport via facilitation has been proposed~\cite{valencia2023rydberg}.

\let\oldaddcontentsline\addcontentsline

\renewcommand{\addcontentsline}[3]{}

\medskip
\begin{acknowledgments}
\prlsection{Acknowledgements} We thank Leong-Chuan Kwek, Wenhui Li, Francesco Perciavalle, Enrico Domanti, Wayne J. Chetcuti, Davide Rossini and 
Thibault Vogt for discussions. The Julian Schwinger Foundation grant JSF-18-12-0011 is acknowledged. OM and AL also acknowledge support by the H2020 ITN ``MOQS" (grant agreement number 955479) and MUR (Ministero dell’Università e della Ricerca) through the PNRR MUR project PE0000023-NQSTI.
\end{acknowledgments} 

\bibliographystyle{unsrt}
\bibliography{library}

\newpage
\onecolumngrid

\section*{Appendix}

\subsection{Experimental Consideration}
\label{Experimental_Considerations}

Considering the experimental creation, we assume the same experimental procedure that is noted in \cite{morsch2018dissipative}. Here $\ce{^{87}Rb}$ atoms are excited from the ground to the Rydberg state via a two-photon transition as they share the same parity \cite{marcuzzi2014effective}. The first of which,  $\ket{5S_{1/2}}\rightarrow\ket{6P_{3/2}}$, a laser with $\Omega_{420}$ excites the atom to an intermediate state. Then from here, another transition occurs, due to a laser with $\Omega_{1013}$, this excites the atom from $\ket{6P_{3/2}}\rightarrow\ket{70S_{1/2}}$, the Rydberg state. Although for the theoretical description we assume a two-level system. In doing so, we denote an effective Rabi frequency given by the equation in \cite{morsch2018dissipative}, 

\begin{equation}
    \Omega = \sqrt{\frac{\Omega_{420}^{2}\Omega_{1013}^{2}}{4\Delta_{6p}^{2}}}
\end{equation}

We chose our simulation parameters in close accordance with experimental work conducted on Rubidium atoms \cite{valado2016experimental, morsch2018dissipative}. We select to use parameters with values: $\tilde{\Omega}/(2\pi) = 0.7\,$MHz, $\tilde{\kappa}/(2\pi) = 2\,$kHz, $\tilde{\gamma}/(2\pi) = 0.7\,$MHz and $\tilde{C}_{6}/(2\pi) = 109\,$GHz$\mu$m$^{6}$. We decide to work in units of $\Omega = 1$ and covert the other parameters accordingly. We convert the dephasing, $\gamma = \tilde{\gamma}/\tilde{\Omega} = 1$, the decay, $\kappa = \tilde{\kappa}/\tilde{\Omega} = 0.003$. For the interaction, we fix the values of $\tilde{\Delta}_{\text{f}} = 7\,$MHz and $\tilde{r}_{\text{f}} = 5\,\mu\text{m}$ therefore $C_{6} =  \tilde{C}_{6}/(\tilde{\Omega} \, \tilde{r}_{\text{f}}) = 10$. We provide these values to a similar range to that of experimental work.

For our systems, we consider detuning with spatial variation. In practice, such conditions can be implemented by suitably shifting the excitation laser frequency (for example through an acousto-optic modulator) and then by exciting specific portions of the Rydberg network with different frequencies.

Note that it is possible to control $\kappa$ in experiment following the scheme presented in~\cite{kazemi2023driven} and experimentally shown in~\cite{begoc2023controlled} via an intermediate level coupled with the Rydberg state.

\subsection{Classical equation vs full simulation}
\label{Classical_vs_Quantum_Simulation_Subsection}

\begin{figure}[htbp]
    \centering
    \includegraphics[width=0.3\textwidth]{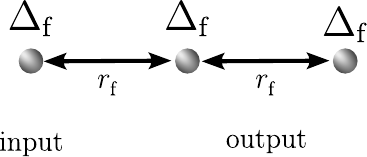}
    \caption{A linear chain of $N = 3$ Rydberg atoms. Positioned at the facilitation radius, $r_{\text{f}}$ between adjacent atoms, the facilitation constraint is fulfilled.} 
    \label{Sche_Transport}
\end{figure} 

\begin{figure}[htbp]
    \centering
    \includegraphics[width=0.9\textwidth]{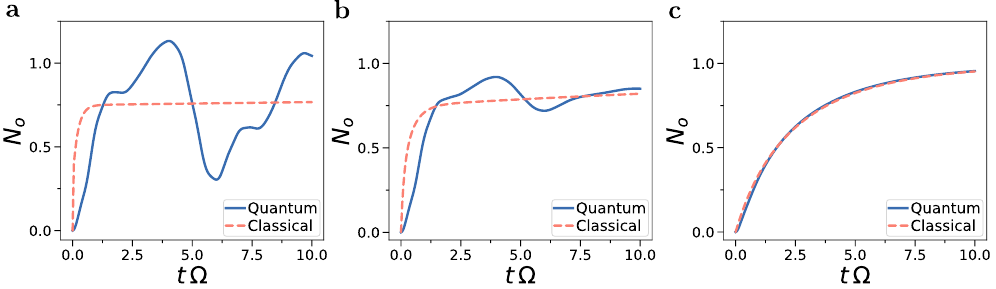}
    \caption{Comparison of the full simulation with the Lindblad master equation and the classical approximation for different values of $\gamma$ with the parameters $N=3$, $C_{6} = 10$ and $\kappa = 0.003$. \idg{a} $\gamma = 0.1$, \idg{b} $\gamma = 1$ and \idg{c} $\gamma = 10$.} 
    \label{C_vs_Q_1D}
\end{figure}

Here we compare the simulation with the full quantum equation 2 against the classical approximation 3, derived for the limit of strong dephasing $\gamma\gg\Omega$. For the system in Fig.~\ref{Sche_Transport}, both the dynamics are represented in Fig.~\ref{C_vs_Q_1D}. We observe that for $\gamma\ge\Omega$, the classical equations are a good approximation to the full quantum dynamic. Beyond this limit, $\gamma < \Omega$, the quantum coherence between the atoms becomes too large, therefore there is a discrepancy between the two evolutions. 

\subsection{Transport}
\label{Transport_sub_section}
We study the transport in a linear chain of Rydberg atoms as shown in Fig.~\ref{Sche_Transport}. We set the inter-atom distance to the facilitation radius $r_{\text{f}}$ and detuning $\Delta_{\text{f}}$. We evolve the system with an initial excitation in the input. 

\begin{figure}[htbp]
    \centering
    \includegraphics[width=0.9\textwidth]{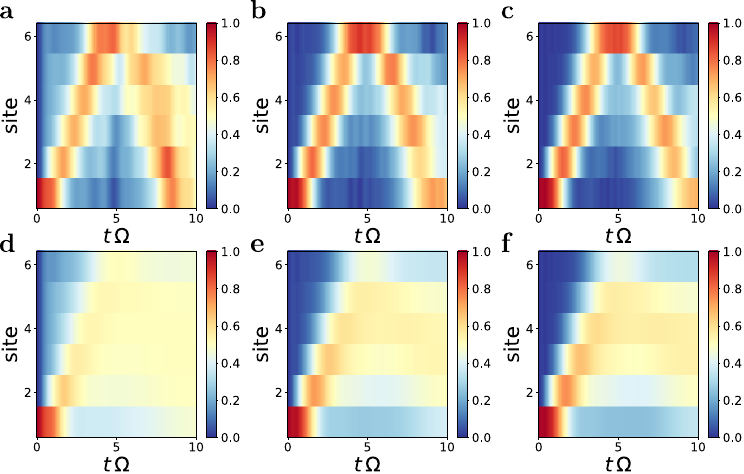}
    \caption{Evolution of the excitation density for an $N = 6$ one-dimensional array of Rydberg atoms with an increasing value of $C_{6}$. Upper panel: $\gamma = 0, \kappa = 0$ and lower panel: $\gamma = 1, \kappa = 0.003$. \idg{a, d} $\text{C}_{6} = 5$ \idg{b, e} $\text{C}_{6} = 10$ \idg{c, f} $\text{C}_{6} = 15$} 
    \label{Transport}
\end{figure}

The dynamics of excitations are shown in Fig.~\ref{Transport}. To observe the full quantum effects, we use the quantum equation and consider two cases: $\gamma = \kappa = 0$ and $\gamma = 1, \kappa = 0.003$.  We observe a propagation of excitation throughout $N = 6$ sites and then a "back-reflection" towards the input. In the situation with zero dephasing and decay, the individual atoms transition between the ground and Rydberg state via the Rabi driving frequency, $\Omega$. The interaction, $C_{6}$, plays an additional role in the correlation in the interaction term, resulting in the excitation being back-reflected at every site. 

With increasing dephasing $\gamma$ and decay $\kappa$, the back reflection is less dominant in the dynamics as the excitation density decreases with increasing propagation distance. We also find that after becoming excited, the atoms are not driven directly to the ground state due to the dephasing destroying the coherence of the state.

\subsection{Diode}
\label{Diode_sub_section}

\begin{figure}[htbp]
    \centering
    \includegraphics[width=0.6\textwidth]{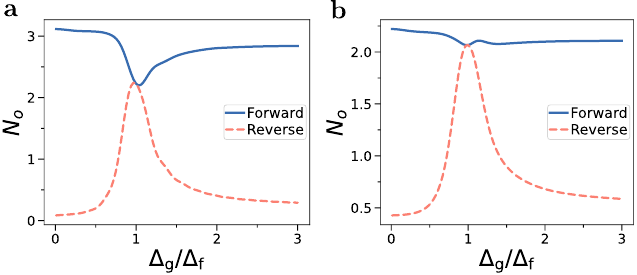}
    \caption{Number of excitations in the output $N_{\text{o}}$ for the forward and reverse direction of the diode for different values of $\Delta_{\text{d}}/\Delta_{\text{f}}$. The parameters are: $N = 6$ and $C_{6} = 10$ \idg{a} $\gamma = 0$, $\kappa = 0$ \idg{b} $\gamma = 1$, $\kappa = 0.003$.} 
    \label{Diode_Plot_Forward_Reverse}
\end{figure}

We now identify a good choice for the gate atom's detuning $\Delta_{\text{g}}$ for the individually trapped diode. We evolve the diode with $N =6$ atoms for $0 < \Delta_{\text{g}}/\Delta_{\text{f}} \le 3$ in both the forward and reverse direction. We measure the number of excitations in the output at $t\Omega = 4.60$ ($t\Omega = 3.20$) with (without) dephasing. We choose these times as here we find the highest number of excitations for the forward direction as studied previously for the switch. We show $N_{\text{o}}$ against $\Delta_{\text{g}}$ in Fig.~\ref{Diode_Plot_Forward_Reverse}. Note that the modified distance $r_{\text{g}}=(-C_6/\Delta_{\text{g}})^{1/6}$ depends on $\Delta_{\text{g}}$. For $\Delta_{\text{g}}\approx\Delta_{\text{f}}$, there is no difference between forward and reverse direction, thus this parameter regime cannot be used for a diode. In contrast, we find a large difference between forward and reverse away from this point. We consider the diode with $\Delta_{\text{g}}/\Delta_{\text{f}} = 2$, whereas values of $\Delta_{\text{g}}/\Delta_{\text{f}} \geq 2$ are sufficient for non-reciprocal behaviour.

\end{document}